\documentclass[twoside]{ae100prg}
\usepackage{graphicx}
\usepackage[breaklinks]{hyperref}
\usepackage{booktabs}

\begin{document}
\title{Quantum Gravity: The view from particle physics}

\author{Hermann Nicolai}

\address{Max Planck Institute for Gravitational Physics\\
(Albert Einstein Institute)\\ Am M\"uhlenberg 1, 14476 Golm,
Germany}

\email{nicolai@aei.mpg.de}

\begin{abstract}
This lecture reviews aspects of and prospects for progress towards 
a theory of quantum gravity from a particle physics perspective, also 
paying attention to recent findings of the LHC experiments at CERN.
\end{abstract}

\section{Introduction}


First of all I would like to thank Ji\v{r}\'{i}  Bi\v{c}\'ak for inviting me to
this prestigious conference in commemoration of Einstein's stay
in Prague a hundred years ago. Although it was only a short stay, as 
Einstein left Prague again after little more than one year, it was
here that he made major progress towards the final version
of General Relativity, and surely the beauty of this city must have
played an important inspirational part in this endeavor. 

In view of the more general nature of this conference, I have decided 
not to give a technical talk on my current work, but rather to
present some thoughts on the state of quantum gravity from the
point of view of a particle physicist, but with an audience of general relativists
in mind. Taking such a  point of view is quite appropriate, as LHC is about 
to end it first phase of experiments, with the solid evidence for a scalar 
boson that has all the requisite properties of a Higgs boson as the 
main outcome so far. This boson was the final missing link in the 
Standard Model of Particle Physics (or SM, for short), and therefore its 
discovery represents the final step in a story that has been unfolding for
almost 50 years. Equally important, as the CERN experiments continue 
to confirm the Standard Model to ever higher precision, with (so far)
no indications of `new physics', it is also a good time to ask whether these 
results can possibly offer any insights into quantum gravity. So my main
message will be that we should not ignore the hints from particle
physics in our search for quantum gravity!

I do not think I need to tell you {\em why} a theory of quantum 
gravity is needed, as some of the key arguments were already
reviewed in other talks at this conference. There is now ample evidence
that both General Relativity (GR) and Quantum Field Theory (QFT)
are incomplete theories, and both are expected to break down
at sufficiently small distances. The generic occurrence 
of space-time singularities in GR is an unavoidable feature 
of the theory, indicating that classical concepts of space and time
must be abandoned at distances of the order of the Planck scale.
Likewise, there are indications of a breakdown of conventional QFT 
in this regime. Accordingly, and in line with the title of this lecture, I would therefore 
like to concentrate on {\em the lessons from particle physics} 
pointing beyond QFT and conventional concepts of space and time.

In its current incarnation, QFT mainly relies on perturbation 
theory. The ultraviolet (UV) divergences that inevitably appear in
higher order Feynman diagrams require a carefully crafted 
procedure for their removal, if one is to arrive at testable
predictions. This renormalization prescription
in essence amounts to an order by order tuning of
a finite number of parameters by infinite factors. Although
mathematically on very shaky grounds, this procedure  
has produced results in stunning agreement
with experimental findings, with a precision unmatched by 
any other scheme in the physical sciences. The most famous
example is, of course, the QED prediction of the anomalous
magnetic moment of the electron, but the agreement between
very recent precision measurements at LHC and the theoretical predictions 
of the Standard Model is now equally impressive. Yet,
in spite of this extraordinary success there is good reason
to believe that neither the SM in its present form nor any of its quantum field 
theoretic extensions (such as the supersymmetric versions of the SM)
are likely to exist in a strict mathematical sense. 
The ineluctable conclusion therefore seems to be that the UV completion 
of the SM requires something beyond QFT as we know it. 

The difficulties in both GR and conventional QFT
have a common origin. In both frameworks space-time is 
assumed to be a {\em continuum}, that is, a differentiable manifold. 
As a consequence, there should exist no obstacle of 
principle in going to arbitrarily small 
distances if either of these theories were universally valid.
Nevertheless, the very nature of quantum mechanics
suggests that its principles should ultimately also apply 
to space-time itself, whence one would expect the emergence
of a grainy structure at the Planck scale. Indeed, and in
spite of their disagreements, almost all approaches 
to quantum gravity\footnote{With the possible exception
 of the Asymptotic Safety program \cite{Weinberg}.}
are united in their expectation that something 
dramatic must happen to space-time at Planck scale distances, where
the continuum should thus give way to some kind of {\em discretuum}. 

A second, and related, source of difficulties is the assumption
that elementary particles are to be treated as {\em point-like} 
excitations. And indeed, there is not a shred of a hint so far
that would point to an extended structure of the fundamental constituents of 
matter (quarks, leptons and gauge bosons), so this assumption seems well
supported by experimental facts. Nevertheless, it is
at the root of the ultraviolet infinities in QFT. 
Moreover, it is very hard to do away with, because the 
point-likeness of particles and their interactions seems to be required by 
both relativistic invariance and locality/causality -- building
a (quantum) theory of relativistic extended objects 
is not an easy task!  In  classical GR, the 
very notion of a point-particle is problematic as well, 
because any exactly point-like mass would have to be a 
mini black hole surrounded by a tiny horizon, and thus 
the putative point particle at the center would move on 
a space-like rather than a time-like trajectory. Again, one is
led to the conclusion that these concepts must be replaced by
more suitable ones in order to resolve the inconsistencies 
of GR and QFT. \\[2mm]
Current approaches to quantum gravity can be roughly put into
one of the two following categories (for a general overviews 
see e.g. \cite{Einstein,Kiefer,100}\footnote{As there is a vast literature
  on this subject, I here take the liberty of citing only a few representative
  introductory texts, where more references can be found.}). 

\begin{itemize} 
\item According to the first hypothesis quantum gravity in essence 
is nothing but the {\em non-perturbative} quantization of Einstein Gravity 
(in metric/connection/loop or discrete formalism). Thus GR, suitably 
treated and eventually complemented by the Standard Model of 
Particle Physics or one of its possible extensions, should correctly describe
the physical degrees of freedom also at the very smallest distances.
The first attempt of quantizing gravity relied on canonical quantization,
with the spatial metric components and their conjugate momenta as
the canonical variables, and the Wheeler-DeWitt equation governing the 
dynamics \cite{Kiefer}. Superimposing Schr\"odinger-type wave mechanics on
classical GR, this scheme was still rather close to classical concepts
of space and time. By contrast, modern versions of this approach look 
quite different, even though their starting point
is still the standard Einstein-Hilbert action in four dimensions: 
for instance, the discrete structure that emerges from the loop quantum 
gravity program relies on holonomies and fluxes as the basic
variables, leading to a discretuum made of  spin networks or spin 
foams \cite{Rovelli,Thiemann}.

\item According to the opposite hypothesis (most prominently represented
by string theory \cite{GSW,Polchinski,BLT}) GR is merely an effective 
(low energy) theory arising at large distances from a more fundamental 
Planck scale theory whose basic degrees of freedom and whose dynamics 
are very different from either GR or conventional QFT, and as yet unknown. 
In this view, classical geometry and space-time itself, as well 
as all matter degrees of freedom  are assumed to be `emergent', 
in analogy with the emergence of classical macroscopic physics 
from the completely different quantum world of atoms and molecules.
Likewise, concepts such as general covariance and even background
independence might only emerge in the large distance limit and 
not necessarily be features of the underlying theory. Consequently, 
attempts to unravel the quantum structure of space and time by directly
quantizing Einstein's theory would seem as futile as trying 
to derive microscopic physics by applying canonical quantization 
procedures to, say, the Navier-Stokes  equation. The fundamental reality
might then be something like the abstract space of all conformal field 
theories, only a small subset of which would admit a geometrical
interpretation. The occasional `condensation' of a classical space-time 
out of this pre-geometrical framework would then appear as a rare event.
\end{itemize}

\begin{figure}
\begin{center}
\includegraphics[width=10cm,height=8cm]{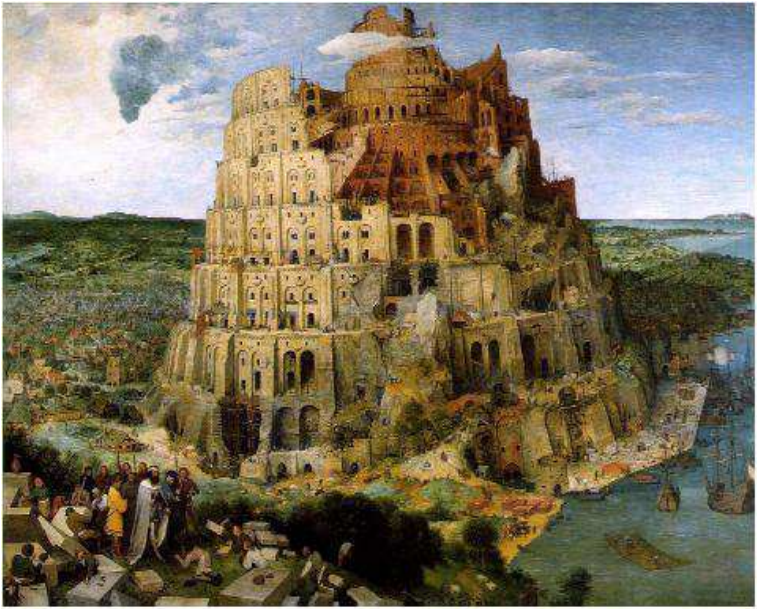}
\caption{The steady progress of Quantum Gravity?} \label{fig1}
\end{center}
\end{figure}

Pursuing different and independent ideas is certainly 
a good strategy as long as we do not know the final answer, 
but it is a bit worrisome (at least to me) that the proponents 
of the different approaches not only base their approaches on 
very different assumptions, but continue to speak languages
that are foreign to one another. Surely, when zeroing in on
the `correct' theory there should be {\em a convergence of ideas
and concepts:} when  Schr\"odinger proposed wave mechanics 
and Heisenberg formulated matrix mechanics, these were
initially regarded as very different, but it did not take long before
it became clear that they were just equivalent descriptions
of the {\em same} theory. Unfortunately, at this time there is no such 
convergence in existing approaches to quantum gravity --
a sign that we are probably still very far from the correct answer!
So let us hope that our noble search will not end like the 
historic event in the Breughel painting shown in Fig.~\ref{fig1}.

\section{The divergence problem}
From the point of view of perturbative QFT 
the basic difference between gravity and matter interactions 
is the non-renormalizability of perturbatively treated GR.
For instance, at two loops the Einstein-Hilbert action must be 
supplemented by the following counterterm cubic in
the Weyl tensor \cite{GS,VdV}~\footnote{There is no need here to distinguish
  between the Riemann tensor and the Weyl tensor, as all terms
  containing the Ricci scalar or  the Ricci tensor can be absorbed
  into (possibly divergent) redefinitions of the metric.}
\begin{equation}\label{Div}
\Gamma^{(2)}_{div}=\frac1\varepsilon\frac{209}{2880}
\frac1{(16\pi^2)^2}\int dVC_{\mu\nu\rho\sigma}
C^{\rho\sigma\lambda\tau} C_{\lambda\tau}^{~~~\mu\nu}\,,
\end{equation}
if the calculations are to produce  {\em finite} predictions
for graviton scattering at this order (the parameter $\varepsilon$
here is the deviation from four dimensions in dimensional
regularization, and must be taken to zero at the end of the calculation).
At higher orders there will
arise similar infinities that likewise must be cancelled
by counterterms of higher and higher order in the Riemann
tensor. Because one thus has to introduce an unlimited
number of counterterms in order to make predictions at arbitrary 
loop orders and therefore has to fix an infinite number of
coupling constants, the theory looses all predictive power.

From the non-renormalizability of perturbatively quantized 
gravity, one can draw quite different conclusions, in particular 
reflecting the two opposite points of view cited above. According to the 
string/supergravity `philosophy', a consistent quantization of 
gravity necessarily requires a modification of Einstein's 
theory at short distances, in order to cancel the infinities. 
This entails the necessity of (possibly supersymmetric) matter 
and in particular fermions, thus furnishing a possible {\em raison 
d'\^{e}tre} for the existence of matter in the world. It was originally
thought that the UV finiteness requirements might single out the unique
maximally supersymmetric field theory -- {\em maximal $N\!=\!8$ supergravity} --
as the prime candidate for a unified theory of quantum gravity, but
that theory was eventually abandoned in favor of superstring 
theory as it became clear that maximal supersymmetry
by itself may not suffice to rule out all possible counterterms.
Superstring theory gets rid of the divergences in a different way, 
by resolving the point-like interactions of QFT into extended 
vertices, relying not only on supersymmetry, but also
on a specifically `stringy' symmetry, {\em modular 
invariance}. Nevertheless, very recent developments \cite{Bern} have 
rekindled the debate whether $N=8$ supergravity could, after all, be a
purely field theoretic extension of Einstein's theory that is 
UV finite to all orders.

On the other side, one can argue that the UV divergences 
of perturbative quantum gravity are merely an artifact of the 
perturbative treatment, and will disappear upon a  proper 
{\em non-perturbative quantization} of Einstein's theory. In this view, perturbative
quantization is tantamount to `steamrollering  Einstein's beautiful 
theory into flatness and linearity' (R.~Penrose):  by giving up
the core features of GR, namely general covariance and background 
independence, one cannot expect to get any sensible complete 
answer. This is the point of view adopted by most of the `non-string' 
approaches, see e.g. \cite{Rovelli,Thiemann}. The concrete 
technical implementation of this proposal invokes unusual properties which
are very different from familiar QFT concepts; for instance, the  
finiteness properties of canonical loop quantum gravity hinge 
on the non-separability of the kinematical Hilbert space.\footnote{The
   non-separability of the kinematical Hilbert space is also a crucial ingredient
   in proposals to resolve 
   space-time singularities in loop quantum cosmology \cite{Bojo}.}
These features are at the origin of the difficulties that
this approach encounters in recovering a proper semi-classical limit, 
and make it difficult to link up with established QFT results. Also for 
this reason there is so far no clue from non-perturbative quantization 
techniques as to what the detailed mechanism is that could 
dispose of the divergence (\ref{Div}).

There is a third (and more conservative) possibility that has lately received 
considerable attention, namely {\em asymptotic safety} \cite{Weinberg,RS}.
This is the proposal that the non-renormalizability of quantum 
gravity can be resolved by a kind of {\em non-perturbative renormalizability}, 
in the sense that there might exist a non-trivial fixed point to 
which the theory flows in the UV. Such a behavior would be 
similar to QCD, which flows to an asymptotically free theory 
in the UV, but the UV fixed point action for Einstein's theory would not be free, 
but rather characterized by higher order contributions in the Riemann 
tensor. In this case there would be no such thing as a `smallest distance', 
and space-time would remain a continuum below the Planck scale. 
If it works, asymptotic safety is probably the only way to tame
the divergences of perturbatively quantized gravity {\em without} 
resorting to the cancellation mechanisms invoked by supergravity 
and superstring theory. The hypothetical non-perturbative 
renormalizability of gravity would also have to come to the 
rescue to resolve the inconsistencies of standard QFT.

However, independently of which point of view one prefers, 
it should be clear that {\em no approach to quantum gravity
can claim complete success that does not explain in full and 
convincing detail the ultimate fate of the divergences of
perturbative quantum gravity.}

\section{The role of matter}

A main point of disagreement between 
the different approaches concerns the role of matter degrees
of freedom. At least up to now, in modern loop and spin foam quantum 
gravity or other discrete approaches `matter does not matter', in the 
sense that matter degrees of freedom are usually treated 
as more of an accessory that can be added at will once 
the quantization of pure gravity has been achieved.
By contrast, to a supergravity/string practitioner the matter content
of the world must play a key role in the search for quantum gravity.
String theory goes even further, positing that the graviton 
is but one excitation (although a very distinguished one)
among an infinite tower of quantized vibrational modes that
should also include all the constituents of matter, and that 
all these degrees of freedom are required for the consistency 
of the theory.

Perhaps it is fitting at this point to recall what Einstein 
himself remarked on the different character of the two sides of 
his field equations: the left hand side is pure geometry 
and beautifully unique, thus made of marble, whereas the right 
hand side has no share in this beauty:
\begin{equation}
\underbrace{R_{\mu\nu}-\frac12g_{\mu\nu}R}_{\rm
Marble}=\underbrace{\kappa T_{\mu\nu}}_{\rm Timber?}
\end{equation}
Indeed, the question that occupied Einstein until the end of his
life was this:  can we understand the right hand side
geometrically, thereby removing its arbitrariness? Put
differently, is there a way of massaging the right hand side 
and moving it to the left hand side, in such a way that
everything can be understood as coming from some sort of 
generalized geometry?

Over the last ninety years there has been some remarkable progress in 
this direction (see e.g. the reprint volume \cite{ACF}), but we still do 
not know whether these ideas really pan out. Already in 1921, T. Kaluza 
noticed that electromagnetism (Maxwell's theory) can be understood
as originating from a five-dimensional theory of pure gravity;
later O. Klein extended this proposal to non-abelian 
gauge interactions. The idea of higher dimensions and of finding 
a geometrical explanation for the existence of matter continues 
to hold fascination to this day, most recently with the idea of
{\em large} extra dimensions (whereas the original 
Kaluza-Klein proposal assumed the extra dimensions  
to be of Planck size in extension). Supersymmetry and 
supergravity may likewise be viewed as variants of the 
Kaluza-Klein program: they generalize ordinary geometry 
by including {\em fermionic} dimensions. This leads to the 
replacement of ordinary space-time by a superspace consisting 
of bosonic (even)  and fermionic (odd) coordinates, thus 
incorporating {\em fermionic matter} into the geometry \cite{BW}. 
Accordingly, the possible discovery of supersymmetric particles 
at LHC could be interpreted as evidence of new dimensions 
of space and time.

There is not so much discussion of such ideas in the `non-string' context,
where neither unification nor extra dimensions feature prominently
(the Ashtekar variables exist only in three and four space-time 
dimensions), and the focus is on Einstein gravity in four dimensions.
Although loop and spin foam quantum gravity are thus  very much tuned to
four dimensions, there have nevertheless been attempts to extend
the framework to higher dimensions, specifically by replacing
the groups $SU(2)$ (for loop quantum gravity) or  $SO(4)$ or 
$SO(3,1)$ (for spin foam models) by bigger groups, with higher
dimensional analogues of the Ashtekar variables, but it is not clear 
whether one can arrive in this way at a unification 
properly incorporating the SM degrees of freedom.

\section{The hierarchy problem}

A problem that is not so much in the focus of the GR
community, but much discussed in the particle physics community
concerns the question of scales and hierarchies. The gravitational force
is much weaker than the other forces (as one can see immediately by
comparing the gravitational attraction between the electron
and the nucleus in an atom with the electric Coulomb force, which 
differ by a factor $10^{-40}$). The so-called hierarchy problem,
then, is the question whether this huge difference in scales can
be `naturally' understood and explained.~\footnote{Of course,
  the biggest and most puzzling hierarchy problem concerns 
  the smallness of the observed cosmological constant.}
In particle physics the problem is reflected in the mass 
hierarchies of elementary particles. Already by itself, the 
observed particle spectrum covers a large range of mass values: light 
neutrinos have masses of less than $1$~eV, the lightest quarks have
masses of a few MeV while the top quark, which is the
heaviest quark discovered so far, has a mass of around $173$~GeV,
so even quark masses differ by factors on the order of $10^5$,
presenting a `little hierarchy problem'. 
But all these mass values are still tiny in comparison
with the Planck scale, which is at $10^{19}$~GeV! This, then, is
the distance that theory has to bridge: at the lower end it is the
electroweak scale that is now being explored at LHC, while at the
higher end it is Planck scale quantum gravity.

A much advertized, but very QFT specific, indication of the problem is the 
occurrence  of quadratic divergences in radiative corrections to the scalar 
 (Higgs) boson mass, which require an enormous fine-tuning 
to keep the observed value so small in comparison with the Planck 
mass, the `natural' value. The absence of quadratic divergences
in supersymmetric theories, where divergences are at most logarithmic, 
is widely considered as a strong argument for low energy 
supersymmetry, and the prediction that each SM particle should 
be accompanied by a supersymmetric partner.

The smallness of the gravitational coupling in comparison
with the other couplings in nature is the main obstacle
towards the verification or falsification of any proposed model of 
quantum gravity. Unless there is a dramatic evolution of the 
strength of the gravitational coupling over experimentally 
reachable energy scales there is no hope of `seeing' quantum
gravity effects in the laboratory. So one needs to find ways and means 
to reason indirectly in order to identify low energy hints of 
Planck scale physics. One possibility might be to look for 
signatures of quantum gravity in the detailed structure of CMB 
fluctuations. The other possibility (which is more in line with this talk) 
is to try to read the signs and hints from the observed structure of the 
low energy world. In the final consequence, this would require
a more or less unique prediction for low energy physics and the 
observed matter content of the world.\footnote{This option is not very
  popular with aficionados of the multiverse or the anthropic principle but,
  interestingly, the hope for a {\em unique} path from quantum gravity 
  to the SM is also prominently visible in the very first papers on the
 heterotic superstring \cite{GHMR,CHSW}.} A more exotic possibility, 
advocated by proponents of large extra dimension scenarios, 
could be an (enormous) increase in the gravitational coupling
strength in the TeV range that would make quantum gravity and 
quantum string effects directly accessible to experiment
(`TeV scale quantum gravity').

At any rate, {\em it remains a key challenge for any proposed theory of 
quantum gravity to offer quantifiable criteria for its confirmation or falsification.}
And the emphasis here is on `quantitative', not on qualitative features that 
might be shared by very different approaches and thus may not suffice to
discriminate between them (for instance, I would suspect this to be the 
case for specific properties of the CMB fluctuations, which may not
contain enough information for us to `read off' quantum gravity). So the 
challenge is to come up with criteria that allow to {\em unambiguously 
discriminate a given proposal against alternative ones}!

\section{From the Standard Model to the Planck scale}

By now, the SM of particle physics is an
extremely well tested theory. It is based on the (Yang-Mills) gauge 
principle with Yang-Mills group $G_{SM}=SU(3)_c\times
SU(2)_w\times U(1)_Y$. Forces are mediated by spin-one gauge
bosons. Matter is made up of spin-$\frac12$ fermions: at this time,
we know of $48$ fundamental fermions which are grouped into three
generations (families) of 16 quarks and leptons each (including
right-chiral neutrinos). There is no evidence so far from LHC 
of any new fundamental fermions.

However, after decades of theoretical research, we still do 
not know what distinguishes the SM gauge group from other 
possible choices. Apart from anomaly cancellations (see below),
the same ignorance prevails with regard to the observed matter
content of the SM. Why does Nature repeat itself with 
three generations of quarks and leptons (another fact 
confirmed by CERN experiments, as well as cosmological 
observations)? What causes symmetry breaking
and what is the origin of mass? And, returning to the question 
of hierarchies, what keeps the electroweak scale stable with 
regard to the Planck scale? (More on this below...)  And,
finally, why do we live in four space-time dimensions?
To tackle these questions, numerous proposals have been put
forward  for physics beyond the Standard Model (or `BSM physics',
for short): Grand Unification (or GUTs, for short), technicolor, 
low energy supersymmetry, large extra dimensions,
TeV scale gravity, excited gauge bosons, and so on.

The main recent progress is the discovery of a scalar
boson by LHC and the strong evidence that this particle has
all the requisite properties of the Higgs boson, especially with the most
recent data indicating that it has indeed spin zero and even parity
(a remaining uncertainty concerns the coupling to the SM fermions, 
which must be proportional to their masses). 
In addition the symmetry breaking mechanism giving 
mass to gauge bosons (`Brout-Englert-Higgs mechanism') has
now been confirmed. However, much to the dismay of many
of my colleagues, no signs of `new physics' have shown up so 
far at LHC. It is therefore not excluded that there may be nothing
more than the SM, augmented by right-chiral  neutrinos, right up to the Planck 
scale, a scenario that is usually referred to as the `Grand Desert'.

\begin{figure}
\begin{center}
\includegraphics[width=10cm,height=8cm]{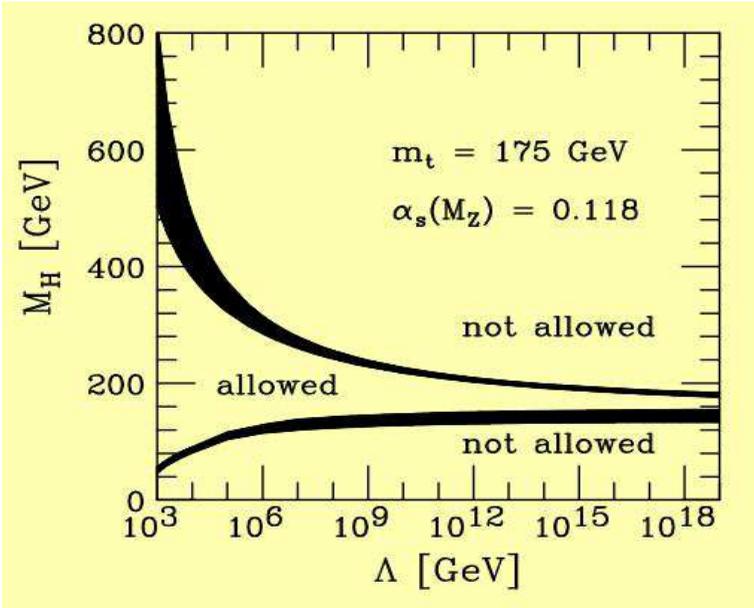}
\caption{Can the SM survive up to the Planck scale? The upper envelope
enforces avoidance of Landau pole for the scalar self-coupling, while the
lower envelope ensures avoidance of vacuum instability \cite{HR} (with 
 an assumed top quark mass of 175 GeV this plot is not quite up to date,
 but this does not affect our main conclusions).  
}
\label{fig2}
\end{center}
\end{figure}

Since the Higgs boson is partly responsible\footnote{Only partly,
  as for instance the  larger part of the proton mass is due to non-perturbative 
  QCD effects!} for the generation of mass, and mass measures
the strength of gravitational coupling, one can reasonably
ask whether these data contain indications of  Planck scale 
physics reaching down to the electroweak scale.
It is here that the question of stability of the electroweak scale 
comes into play. Actually, the stability of the Standard Model is under menace 
from two sides. Following the RG evolution of the scalar self-coupling 
up to the Planck scale, one danger is
the Landau pole where the scalar self-coupling diverges
(as happens for IR free theories with scalar fields)
and the theory  breaks down. The other danger 
is the potential instability caused by the negative 
contribution to the effective potential from the top quark
(the effective potential includes perturbatively computable
quantum corrections to the classical potential).
In Homer's tale, Ulysses has to maneuver his ship between
two formidable obstacles, {\em Skylla} and {\em Charybdis}:
on the one side he must steer it away from 
the rock against which it will crash, and on the 
other side must avoid the sea monster that will swallow the ship. 
The plot of figure \ref{fig2} illustrates the situation.
The vertical direction is the Higgs mass, which grows proportionally
with the Higgs self-coupling. Through the RG evolution
the scalar self-coupling will eventually hit the Landau pole,
at which point the theory crashes. Exactly where this happens
depends very delicately on the Higgs mass. The upper 
curve (the rock) in figure \ref{fig2} \cite{HR} relates
the location of the Landau pole directly to the Higgs mass.
The lower curve (the sea monster) is the constraint from the
negative contribution to the effective Higgs potential. If this
contribution is too negative, it will make the potential unbounded
from below. For the SM, the negative contribution is mainly due to the 
top quark, and it is a danger precisely because the top quark mass is so
large. As a result, if you want to salvage the SM up 
to the Planck scale, there remains only a very narrow strip for the
SM parameters (masses and couplings). The recent results and data from 
LHC indicate that Nature might indeed avail itself of this possibility:
with a Higgs mass of about 125 GeV, the Landau pole can 
safely hide behind the Planck scale, but this value is so low 
that the  SM hovers on the brink of instability! See also \cite{SW}
for an interesting interpretation of this value from the point of view
of asymptotic safety.

To be sure, the potential instability of the effective potential 
is the worse of the two dangers. Namely, the occurrence of a Landau 
pole can always be interpreted as signalling the onset of 
`new physics'  where new degrees of freedom open up and
thereby cure the problem. A well known example of this phenomenon 
is the old Fermi theory of weak interactions, where the non-renormalizable 
four-fermion vertex is dissolved at sufficiently high energies
by new degrees of freedom ($W$ and $Z$ bosons) 
into a renormalizable and unitary theory. Another example would
be the (still conjectural) appearance of supersymmetry 
in the TeV range, which would remove
the Landau pole and also ensure full stability, as the effective
potential in a globally supersymmetric theory is always 
bounded from below (this is no longer true for local supersymmetry).

If we find out whether or not there are genuine new degrees of freedom 
in the TeV range of energies, we may also get closer to answering
the old question of the ultimate divisibility of matter, namely the 
question whether the known particles possess further substructures,
sub-substructures, and so on, as we probe smaller and smaller
distances. Translated into the UV, the question can be rephrased 
as the question whether there are any `screens' ( $\equiv$ scales of `new physics') 
between the electroweak scale and the Planck scale. The more of 
such screens there were  between the electroweak scale and the Planck scale, 
the less one would be be able to `see'  of Planck scale physics. On the 
other hand, the fewer there are, the harder becomes the challenge 
of explaining low energy physics from Planck scale physics.  

\begin{figure}
\begin{center}
\includegraphics[width=10cm,height=8cm]{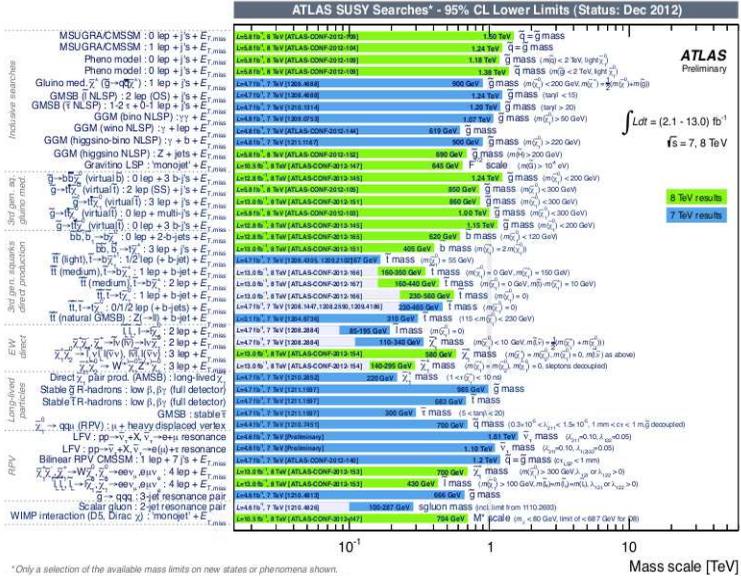}
\caption{Low energy supersymmetry?} \label{fig3}
\end{center}
\end{figure}

\begin{figure}
\begin{center}
\includegraphics[width=10cm,height=8cm]{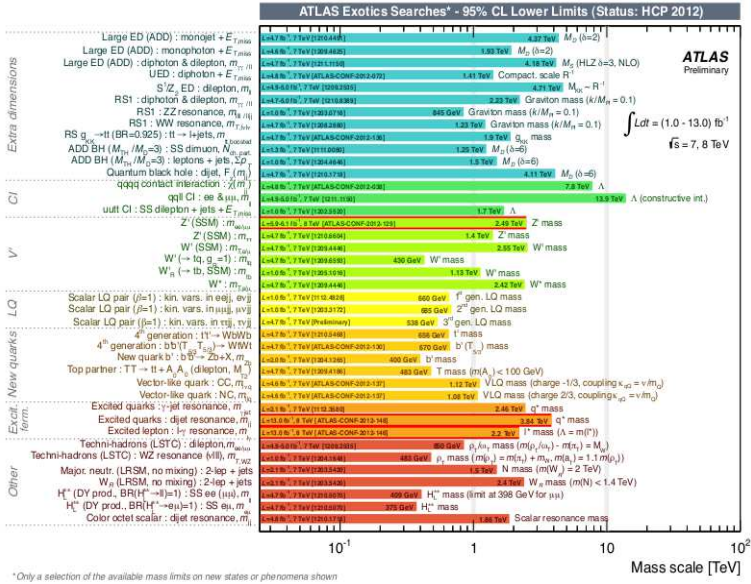}
\caption{Low energy exotics?} \label{fig3b}
\end{center}
\end{figure}

LHC is now testing a large number of `BSM' proposals, and actually 
eliminating many of them.\footnote{The two plots shown below have
 been downloaded from the CERN website
{\tt https://twiki.cern.ch/twiki/bin/view/AtlasPublic/CombinedSummaryPlots}
 where also a summary of many further results can be found.}
Figure \ref{fig3} shows the latest exclusion plot 
from December 2012 on the search for various signatures of supersymmetry.
Figure \ref{fig3b} shows a similar plot from December 2012 for `exotica'
such as large extra dimensions, mini black holes, excited $W$ and $Z$
bosons, quark substructure, and so on, with some exclusions 
already reaching up to 10 TeV. 
As you can see, even to refute only a representative subset
of the proposals on the market and  to keep up with the flood 
of theoretical ideas is a painstaking effort for the experimentalists,
requiring teams of thousands of people and thousands of computers! 

There are theoretical indications that  LHC may not reveal much
new beyond the SM Higgs boson, and thus {\em no} screens between the 
electroweak scale and the Planck scale. It is a remarkable fact that 
the SM Lagrangian is {\em classically conformally invariant
except for a single term}, the explicit mass term in the Higgs potential.
But in a classically conformal theory mass terms can in principle 
be generated by the conformal anomaly, the quantum mechanical 
breaking of conformal invariance, which could also trigger spontaneous
symmetry breaking \cite{CW}. Maybe no explicit mass terms are needed
in the SM Lagrangian, and the mechanism stabilizing the electroweak scale
is {\em conformal symmetry} rather than low energy supersymmetry 
\cite{Bardeen,MN}? A further hint in this direction comes from
the flows of the SM couplings under the renormalization group:
it almost looks like these couplings could `keep each other under control' 
so as to prevent both Landau poles and instabilities right up to the 
Planck scale! This is because bosons and fermions contribute 
with opposite signs to the corresponding $\beta$-functions.
The scalar self-coupling would normally blow up under the flow, 
but is kept under control by the top quark contribution which delays the 
appearance of the Landau pole until after the Planck scale.
The same mechanism is at work for the top quark (Yukawa) coupling
which is asymptotically not free either: it, too, would blow up,
but is kept under control by the strong coupling $\alpha_s$, again
shifting the Landau pole beyond the Planck scale. Finally $\alpha_s$ 
itself is kept under control in the UV by asymptotic freedom. 

In summary, it could just be that {\em the mass patterns and the couplings in the 
Standard Model precisely conspire to make the theory survive to the Planck scale.} 
In this case there would be no `new physics' beyond the electroweak scale 
and the theory would have to be embedded directly {\em as is}
into a Planck scale theory of quantum gravity. In my opinion, this may
actually be our best chance to gain direct access to the Planck scale,
both theoretically and experimentally!


\section{Anomalies}

There is another remarkable property of the SM which may be 
interpreted as a hint of how Planck scale physics could affect 
low energy physics, and this is the complete cancellation of gauge 
anomalies (see \cite{Bertlmann} for an introduction and many
references to the original work). Anomalies 
occur generically when a classical Lagrangian is invariant 
a symmetry, but that symmetry cannot be preserved by the regularization
that quantization requires. When the regulator is removed there is a finite
remnant, and this is referred to as the anomaly, an ${\cal O}(\hbar)$ violation 
of a classical conservation laws. The classic example of such a 
symmetry is chiral invariance  that explains the 
(near-)masslessness of fermions, but cannot be regulated, 
leading to the famous axial anomaly in QED that accounts 
for the decay of the $\pi^0$ meson. 

When anomalous currents are coupled to gauge fields, the
anomaly can deal a fatal blow to the theory.
Recall that the coupling of a gauge field $A_\mu$ to charged
matter generally takes the Noether form $\propto A_\mu J^\mu$, where
$J^\mu$ is the classically conserved matter current. In the presence
of an anomaly the variation of this term would give
\begin{equation}
\int \delta A_\mu J^\mu= \int \partial_\mu \omega J^\mu = -\int \omega\partial_\mu
J^\mu \propto {\cal O}(\hbar) \neq0\,.\label{eq1}
\end{equation}
Gauge invariance would thus no longer hold, and this
violation would destroy the renormalizability of the SM
and thereby its predictivity. To verify that all gauge anomalies 
and gravitational anomalies cancel in the Standard Model
requires the computation of various triangle diagrams
with chiral fermions circulating in the loop, and involves traces
of the form ${\rm Tr}\, T^a \{ T^b , T^c\}$, where $T^a$ belong to the
Lie algebra of the Standard Model gauge group.
More specifically, the calculation reduces to the evaluation of
\begin{equation} \label{TTT}
\sum \pm \, {\rm Tr\,} YYY = 
\sum \pm \, {\rm Tr\,} tt Y = \sum \pm \, {\rm Tr\,} Y = 0 
\end{equation}
where $Y$ is the electroweak hypercharge, and $t$ denotes
any generator of $SU(2)_w$ or $SU(3)_c$; the sum runs
over all SM fermions, with `+' for positive and `$-$' for negative chirality
fermions. If you work through  the whole list of such diagrams you will find that they 
all `miraculously' sum up to zero \cite{Bertlmann}. From (\ref{TTT}) it is obvious that the 
cancellation would be trivial if the SM were a vector-like theory 
with no preferred handedness or chirality. Remarkably, Nature prefers 
to break parity invariance, and  to do so subtly in a way that 
maintains the renormalizability, hence consistency. In fact,
the anomaly cancellations fix the fermion content almost 
uniquely to what it is, separately for each generation. Therefore, despite its
`messy' appearance {\em the Standard Model is surprisingly unique, 
and also surprisingly economical for what it does!}  

There are two crucial features that must be emphasized here.
The first is that {\em a proper anomaly does not and must 
not depend on how the theory is regulated.}  Secondly,
anomalies are often regarded as a perturbative phenomenon, 
but this is not strictly true. The famous Adler-Bardeen theorem 
asserts that the anomaly is entirely due to the one-loop contribution, 
and that there are thus no further  contributions beyond one loop. 
In other words, the one-loop result is {\em exact to all orders}, hence 
non-perturbative! 

Anomalies should also be expected to play a role in quantum gravity,
and in determining whether a specific proposal is ultimately consistent or not.
For instance, the classical constraint algebra of General Relativity
in the Hamiltonian formulation has the schematic form
\begin{equation}
\{D,D\}\sim D\,,\quad \{D,H\}\sim H\,,\quad \{H,H\}\sim D
\end{equation}
where $D$ and $H$, respectively denote the diffeomorphism 
constraints and the Hamiltonian constraint, and this algebra
is expected to be modified by quantum corrections.
This expectation is borne out by the simplest example, matter-coupled 
quantum gravity in two space-time dimensions. Here the  most general
form of the space-time diffeomorphism algebra including anomalies 
is known to take the form
\begin{equation}
~[T_{\pm\pm}(x),T_{\pm\pm}(y)]=\delta'(x,y)\Big(T_{\pm\pm}(x)
+T_{\pm\pm}(y)\Big)+\hbar c\delta'''(x,y)\,,\label{eq2}
\end{equation}
where $x,y\in\mathbb{R}$, $T_{\pm\pm}:=H\pm D$ and $c$ is the
central charge. As is well known, virtually all of string theory
hinges on the non-zero value of the central charge $c$!
Unfortunately in higher dimensions, there exists neither an analogous
uniqueness result, nor even a classification of what the
anomalies may be. The main difficulty here is that higher-dimensional
 diffeomorphism algebras are `soft', which means that Lie algebra 
 structure `constants' are not really constant, but field dependent.

\section{Outlook}

So where do we stand? At this time there is a
growing array of proposals for quantum gravity, based on a 
variety of different and even mutually contradictory
assumptions and hypotheses. The following is a selection 
of current approaches (to which you may add
your own favorite):

\begin{itemize}
\item Supergravity, Superstrings and M-Theory

\item AdS/CFT and Holography

\item Path integrals: Euclidean, Lorentzian, matrix models, ...

\item Canonical Quantization (metric formalism)

\item Loop Quantum Gravity (with either connections or holonomies)

\item Discrete Quantum Gravity: Regge calculus, (causal) dynamical triangulations

\item Discrete Quantum Gravity: spin foams, group field theory

\item Non-commutative geometry and space-time

\item Asymptotic Safety and RG Fixed Points

\item Emergent (quantum) gravity from thermodynamics

\item Causal Sets

\item Cellular Automata (`computing quantum space-time')
\end{itemize}

Among these string theory remains the leading contender, not least
because it naturally incorporates (and even requires) matter degrees 
of freedom. Nevertheless, we still do not  have a single hint from experiment 
and observation (for instance, in the form of supersymmetric partners 
to the known elementary particles) that it is indeed the right theory. 
Perhaps it is thus not so surprising that `non-string approaches' 
have been gaining in popularity over the past few years.  

Having grown out of particle physics and being modeled on its 
basic concepts, string theory has no problem of principle in connecting 
to low energy physics; being a perturbative approach, it also has no
difficulties in reproducing the correct semi-classical limit and the 
Einstein field equations. But  after more than two decades of effort, 
string theory is still struggling to reproduce the Standard Model {\it as is},  
that is, without the heavy extra baggage that comes with (for 
instance) the supersymmetric extensions of the SM referred to as 
`MSSM', `CMSSM'  or `NMSSM', and so on. Moreover, it has 
considerable difficulties in incorporating a {\em positive} cosmological 
constant -- in fact, like supergravity, superstring theory has an overwhelming preference for
negative $\Lambda $! String theory, as originally formulated, is a 
background dependent and perturbative theory. However, there 
have been important advances and recent developments, especially in connection 
with the AdS/CFT correspondence and gauge/gravity or weak/strong 
dualities, that transcend perturbation theory and have provided
important insights into the non-perturbative functioning of the theory
(see e.g. \cite{BT} for a recent update). Nevertheless, in its present form 
string theory does not offer a convincing  scenario for the resolution of 
(cosmological) space-time singularities, and so far cannot tell us 
what really `happens' to space-time at the Planck scale. 

I have already mentioned the impressive recent advances 
in perturbative QFT techniques \cite{Bern}, yielding evidence
that $N=8$ supergravity may be finite to all orders, contrary to
expectations held for more than 30 years. If this theory 
could be shown to be a purely quantum field theoretic 
extension of Einstein's theory {\em without} UV singularities, 
this would partially undermine one of string theory's chief
arguments why QFT must be abandoned. Of course, this would not
relieve us of the task of working towards a {\em non-perturbative}
understanding of physics at the very shortest distances, as the
putative finiteness by itself would not tell us why and how
the space-time continuum is dissolved at the Planck scale.
And even if the theory turned out to be UV finite, many would doubt
whether $N=8$ supergravity has anything to do 
with `real world physics'. Yet, there is a curious coincidence here:
when supersymmetry is completely broken, eight spin-$\frac12$ fermions 
are converted  into Goldstinos in order render the eight gravitinos 
massive, leaving us with 48 spin-$\frac12$ fermions, 
exactly the right number! Most likely a mirage, but
who knows?~\footnote{On this point, see also \cite{NW}.}

In contrast to string theory the non-perturbative approaches put 
the main emphasis on GR concepts from the very beginning, 
to wit, (spatial) background independence and diffeomorphism 
invariance. Following this avenue has led to intriguing new 
ideas and proposals as to what a quantum space-time might 
actually `look like'. Nevertheless, it is hard to see how such ideas 
could ever be put to a real test (other than internal consistency
checks). A main criticism from the point
of view taken here is that these approaches have not 
incorporated  essential insights from particle physics up 
to now, such as the restrictions from anomaly cancellations. 
Furthermore, the ambiguities related to quantization and the incorporation 
of matter couplings have not been resolved in a satisfactory 
fashion in my opinion, and the recovery of the proper semi-classical 
limit remains an outstanding challenge. 

To conclude let me restate my main worry. In one form or another 
the existing approaches to quantum gravity suffer from a very large 
number of ambiguities, so far preventing any kind of prediction 
with which the theory will stand or fall. Even at the risk of sounding 
polemical, I would put this ambiguity at $10^{500}$ (or even more) -- 
in any case a number too large to cut down for any conceivable 
kind of experimental or observational advance.

\begin{itemize}
\item{\bf Superstring theory} predicts the existence of myriads of
 `consistent' vacua, all of which are supposed to be realized 
 somewhere in the multiverse (or `megaverse') -- leading to 
 the conclusion that essentially {\em anything goes} when it comes
 to answering the questions raised at the beginning of section~5 
 (most notably, it is claimed that the multiverse also `solves' the 
 cosmological constant problem).
\item{\bf Loop quantum gravity} and related approaches are 
compatible with many `consistent' Hamiltonians (or spin foam 
models), and with an essentially arbitrary menu of matter fields. 
Even disregarding technical issues such as quantization 
ambiguities, it looks again like almost {\em anything goes}.
Idem for models of lattice and discrete quantum gravity.
\item{\bf Asymptotic Safety} is an assumption that, according to its
proponents, works almost {\em generically} -- that is, independently 
of the specific `initial' conditions for the RG flows, of the matter
content and even the number of space-time dimensions (if that
number is not extremely large), leaving us with numerous 
`consistent' RG flows.
\end{itemize}

In my view the real question is this: if there are all these `consistent' 
(according to your definition) ans\"atze, does Nature simply pick
the `right' answer at random from a huge variety of possibilities,
or are there criteria to narrow down the number of choices? 
Being exposed to many talks from the different `quantum gravity camps'
I am invariably struck by the success stories  I keep hearing, 
and the implicit or explicit claims that `we are almost there'. 
I, for one, would much prefer to hear once in a while that 
something does {\em not} work, and to see some indications of  
{\em inconsistencies} that might enable us to discriminate 
between a  rapidly growing number of diverging ideas on 
quantum gravity \cite{NPZ,AR}.  If, however,  the plethora of theory 
ambiguities were to stay with us I would conclude that our search 
for an ultimate explanation, and with it the search for quantum gravity,
may come to an ignominious end (like in Breughel's painting). 
I cannot imagine that this is what Einstein had in mind during his 
stay in Prague, nor in the later years of his life when he was 
striving to figure out "the old one's tricks" (or, in the original German, 
"dem Alten auf die Schliche kommen").

So let me repeat my main message:
the incompleteness of the Standard Model is one of the 
strongest arguments in favor of quantizing gravity and 
searching for new concepts replacing classical notions of
space and time. The observed features of
SM may contain important hints of its possible UV
completion and Planck scale physics, and these hints should be given
due consideration in the search for a consistent theory of quantum gravity. 

\vspace{4mm}

\noindent{\bf Acknowledgments:} I would like to thank Jianwei Mei for his 
help in turning my talk into a (hopefully) readable text and Krzysztof Meissner
for many enjoyable  and illuminating discussions on the state of the art.


\end{document}